\documentclass[10pt,letterpaper]{article}
\usepackage[top=0.85in,left=1in,footskip=0.75in]{geometry}

\usepackage{amsmath,amssymb}

\usepackage{changepage}

\usepackage[utf8x]{inputenc}

\usepackage{textcomp,marvosym}

\usepackage{cite}

\usepackage{nameref,hyperref}

\usepackage{microtype}
\DisableLigatures[f]{encoding = *, family = * }

\usepackage[table]{xcolor}

\usepackage{array}

\newcolumntype{+}{!{\vrule width 2pt}}

\newlength\savedwidth

\usepackage{setspace} 
\raggedright
\usepackage[aboveskip=1pt,labelfont=bf,labelsep=period,justification=raggedright,singlelinecheck=off]{caption}

\makeatletter
\renewcommand{\@biblabel}[1]{\quad#1.}
\makeatother

\date{}

\usepackage{lastpage,fancyhdr,graphicx}
\usepackage{epstopdf}
 \newcommand{\benum}{\begin{enumerate}}
 \newcommand{\eenum}{\end{enumerate}}
 
 \newcommand{\bite}{\begin{itemize}}
 \newcommand{\eite}{\end{itemize}}
 
 \usepackage{verbatim}
 \usepackage{color}
 
 \def\a{\alpha } \def\b{\beta }   
    
\def\th{\theta }

\begin{document}
\vspace*{0.2in}

\begin{flushleft}
{\Large
\textbf\newline{Inferring Information Flow in Spike-train Data Sets using a Trial-Shuffle Method}
}
\newline
\\
Benjamin L. Walker\textsuperscript{1},
Katherine A. Newhall\textsuperscript{1}*,
\\
\bigskip
\textbf{1} Department of Mathematics, University of North Carolina, Chapel Hill, North Carolina, USA
\\
\bigskip

%

* knewhall@unc.edu

\end{flushleft}
\section*{Abstract}
Understanding information processing in the brain requires the ability to determine the functional connectivity between the different regions of the brain.  We present a method using transfer entropy to extract this flow of information between brain regions from spike-train data commonly obtained in neurological experiments.    Transfer entropy is a statistical measure based in information theory that attempts to quantify the information flow from one process to another, and has been applied to find  connectivity in simulated spike-train data.  Due to statistical error in the estimator, inferring functional connectivity requires a method for determining significance in the transfer entropy values.  We discuss the issues with numerical estimation of transfer entropy and resulting challenges in determining significance before presenting the trial-shuffle method as a viable option.  The trial-shuffle method, for spike-train data that is split into multiple trials, determines significant transfer entropy values independently for each individual pair of neurons by comparing to a created baseline distribution using a rigorous statistical test.  This is in contrast to either  globally comparing all neuron transfer entropy values or comparing pairwise values to a single baseline value.
 In establishing the viability of this method by comparison to several alternative approaches in the literature, we find evidence that preserving the inter-spike-interval timing is important.  
 We then use the trial-shuffle method to investigate information flow within a model network as we vary model parameters. This includes investigating the global flow of information within a connectivity network divided into two well-connected subnetworks, going beyond local transfer of information between pairs of neurons.

\section*{Introduction}

Understanding the flow of information in the brain is a key step in determining how the brain processes information.  Methods for observing this flow of information include 
non-invasive methods such as MEG or calcium imaging that produce a pseudo-continuous image of activity in a brain region~\cite{orlandi2014transfer,vicente2011transfer} but lack high spatial resolution.  Another approach, multi-unit recording, is an invasive method that inserts electrode arrays into the brain to  detect when nearby neurons fire~\cite{green1958simple,humphrey1990extracellular}, producing a detailed measurement of the activity of a relatively small number of neurons.  This data can be processed to produce a list of spike-times for each observed neuron known as spike-train data~\cite{brown2004multiple}.  
Analyzing this spike-train data can provide a window into how the neurons in the brain interact by identifying causal influences instead of just correlations in average activity.

Early methods of analyzing spike-train data include cross-correlation of spike activity and mutual information~\cite{borst1999information,nirenberg2001retinal,brown2004multiple}.  These methods seek to measure the similarity of two spike-trains, under the assumption that a causal relation between them will lead to a higher degree of similarity. However, these methods are symmetric, and so even if they do reveal a strong similarity they cannot indicate what the direction of causality is.  Granger causality is a non-symmetric method, but requires the assumption of linear interaction~\cite{granger1988some,kaminski2001evaluating}.  The desire to reveal the causal relations in data with a non-parametric approach 
 has led to recent investigations into the use of transfer entropy (TE)~\cite{schreiber2000measuring,kaiser2002information,timme2016high}.

Transfer entropy measures information transfer from a source process to a target process by quantifying the improvement in the ability to predict the future of the target process given knowledge of the past of the source process, over just knowledge of the past of the target. Uncorrelated processes will have an exact TE value of 0, and higher TE values generally correspond to greater transfers of information. Transfer entropy is defined on a pair of processes, but through pairwise evaluation can be extended to an entire network of processes such as a social network~\cite{ver2012information} or neuronal network~\cite{gourevitch2007evaluating,vicente2011transfer}.

When computationally estimating TE values, statistical noise and bias in the estimator often causes positive TE values when no causal relationship is present.  
A natural way to determine whether the computed TE value is significant above the noise level is to choose a cutoff that represents the minimum TE value that corresponds to a significant transfer of information.
However, TE values depend on the amount of information present, so neurons with higher firing rates will naturally have higher TE values.
Because of this, an effort must be made to normalize the TE values onto the same scale before they can be compared, which previous papers have attempted~\cite{gourevitch2007evaluating,li2013estimating}.  Our goal is to present a statistical method that circumvents the need to scale TE values before determining significance.

Our analysis is motivated by experimental data collected on ferrets performing an attention task~\cite{sellers2016oscillatory}. This data consists of sets of 64 12-second trials performed consecutively, with recordings taken from two multi-electrode recorders, one each in the frontal lobe and the parietal lobe. Inspired by this, we consider analysis of models where the network is divided into two densely connected clusters with sparse interconnection.  Also, based on the fact that the experimental task has a fixed duration, we create synthetic data consisting of many short trials 
 instead of one long recording as used previously~\cite{tet} and base our analysis on this data structure. Because the data consists of recordings of only a handful of neurons from two distant brain regions, it is unlikely that any two observed neurons will have a direct, anatomical connection. We must instead be able to detect indirect connections between the neurons, where the signal passes through one or more intermediary neurons, if we want to be able to find any flow of information present between the regions (i.e., functional connections).

Prior literature has found the creation of surrogate data to be an effective way to determine the whether transfer entropy values are significant, using methods such as jittering~\cite{nigam2016rich}, shuffling of trials~\cite{vicente2011transfer,lindner2011trentool}, and shuffling of inter-spike-intervals (ISIs)~\cite{rivlin2006local}.
We have found that effective methods for determining statistical significance in transfer entropy values have not been well explored on spike-train data, which poses different challenges than continuous data.
 We propose the following four guiding criteria for a method to determine flow of information:
\benum
\item \label{item:strong} Strong ability to find weak/indirect connections in a network
\item Robust handling of heterogenous networks
\item No human input into the significance determination process
\item No knowledge of underlying network topology required
\eenum
The reasons for these conditions are as follows:  The first condition ensures that not only strong and direct connections are detected, but also weak and indirect connections, important for investigating functional connectivity.  The second addresses the previously mentioned fact that neurons with higher firing rates can have falsely inflated TE values,  and this could lead to false detection if the method does not adequately handle realistic variability in firing rates and coupling strengths.  The third condition, pertaining to the method's design, is that the {method should not involve human decision-making at any point in the process.} With an increased focus on big data in computational neuroscience, human analysis falls flat in terms of scalability. 
Our last condition is motivated by the experimental data which consists only of recorded spike-trains, and not the structural connectivity.

We primarily consider a trial-shuffle method that meets these criteria, based on the trial shuffle for surrogate data introduced in \cite{lindner2011trentool,vicente2011transfer}.  
The trial-shuffle method makes use of the large number of separate trials, typical in experimental data,  
 to independently assess significance in each pair of observed processes by statistically comparing the distribution of TE values to a computed baseline distribution, which we establish by estimating TE values for the same pair of neurons across different trials. This bypasses the problem of normalization \cite{gourevitch2007evaluating} and prevents focus on only the largest TE values within the network.  This baseline would also include any correlations induced by the experimental task, rather than coupled neurons, thus information flow beyond these low level correlations would be detected.  We then consider how the pairwise analysis can be used to test for coarse-grain flow of information between two network regions, and the effects of partial observation where not all neurons are recorded. Furthermore, our method uses neither human intervention (e.g. \cite{li2013estimating,gourevitch2007evaluating}) nor knowledge of the underlying network topology (e.g. \cite{tet}) to determine significance. The latter is especially important as it is a requirement for analyzing experimental data.

The remainder of the paper is organized as follows.  In the methods section, we present our network model, based on the integrate-and-fire neuron model and then describe transfer entropy and the proposed distribution-based comparison method for determining significance.  In the results section, we begin by looking at how TE estimation is influenced by neuron firing rate, coupling strength, trajectory length, and indirect connections. Next, we compare how creating the baseline distribution by shuffling trials, shuffling inter-spike-intervals, jittering, and time-shuffling behave relative to each other and the false positive method which does not fulfill the above criteria.  Then, having established the validity of the trial-shuffle method, we use it to understand how the parameters in the network model affect the ability to detect transfer of information.  Last, we present our discussions on these results.

\section*{Methods}

\subsection*{Neuronal Network Model\label{sec:Model}}

We create synthetic data consisting of sequences of spike-times by numerically simulating networks of integrate-and-fire model neurons.  This allows us to control the network structure and therefore the flow of information.
  Specifically, in a network of $n$ neurons,
the neuronal membrane potentials $V_i$ obey the stochastic differential equations
\begin{equation}
\label{eqIF}
 dV_i = \alpha_i ( \mu_i-V_i) dt + \beta_i dW_i + \sum_{j,k} c_{ji} \delta(t-\tau_{jk}-d_{ji}) dt  \qquad i=1,2,...,n
 \end{equation}
where the $\mu_i$ are the constant driving input,  the $\a_i>0$ are the time-scale coefficient controlling how fast the voltages approach $\mu_i$, the $\b_i>0$ are the noise coefficients modeling variability in the neuron's input, and the $W_i(t)$ are independent standard Brownian motion processes.  In addition to Eq.~\eqref{eqIF} there is a reset mechanism:
whenever a potential $V_i$ reaches the threshold value of $V_T$, the neuron is said to fire a spike and its potential is reset to $V_R$.  The potentials of the other neurons $i$ are altered by a value $c_{ji}$ after a delay $d_{ji}$. The times $\tau_{jk}$ are the prior spike times of neuron $j$.

For our simulations, we use non-dimensional variables and parameters and solve Eq.~\eqref{eqIF}  using a modified Euler-Maruyama method from time $t=0$ up to time $t=t_{\textrm{max}}$.  We take $V_T=1$ and $V_R=0$, and create the spike train by binning the spike times into bins of length 0.001.  Additionally, for  the sake of simplicity, we will reduce the number of parameters by using $\alpha_i = \alpha=1, \beta_i = \beta$,  and $d_{ji} = 0.01$ corresponding to 10 time-bins.   Except where otherwise noted, we additionally use $\beta = 1$.

To construct the network connections, we point out that neuronal networks are generally sparse, 
 so most values of $c_{ji}$ are 0, and all connections emanating from a neuron are either excitatory $(c_{ji}= w_E>0)$ or inhibitory $(c_{ji} = w_I<0)$.    We consider a few classes of network topology. For investigation of basic properties, we consider several results from directed chain networks, with all excitatory connections. A directed chain network simply consists of $n$ neurons with connections from neuron $k$ to neuron $k+1$ for $ k = 1,2,...,n-1$.  We then construct a more realistic model based on experimental data that focuses on the interaction of two regions of the brain.
To model this in our simulated data, we use a directed stochastic block model~\cite{wang1987stochastic} with $2$ communities, which we describe next. In this paper, the neurons are divided equally between these two communities.  We require a large number of connections within each cluster and a small number of connections between the regions. Depending on the parameters of the simulation, the connections between regions may be allowed in one, both, or neither direction. All possible connections, taken to be from neuron $i$ to neuron $j$, are initialized with an independent probability of 
\[ p_{\text{inside}} = \frac{2 k_1}{n-1} \] if the neurons are in the same cluster. For a pair of neurons which are in different regions, the probability is either
\[ p_{A \to B} = \frac{4 k_A}{n^2}  \quad \textrm{or} \quad
 p_{B \to A} = \frac{4 k_B}{n^2} \]
depending on the direction of the connection. The parameter $k_1$ is the average number of outgoing connections a neuron has to other neurons within its region, $k_A$ is the average number of outgoing connections in total from region A to region B, and vice versa for $k_B$. For example, for a two-region network in which there are only connections from region A to region B, one would set $k_A > 0, k_B = 0$.  In this two-region network we randomly assign neurons to be inhibitory with probability 0.25; otherwise the neuron is excitatory.

We also consider the possibility that not all of the neurons are being observed. In this case we have $n_{\text{obs}}$ neurons observed where $n_{\text{obs}} < n$, where the observed neurons are still evenly split between the regions. Spike trains are then only recorded and analyzed for those neurons that were observed. In this paper, $n_{\text{obs}} = n$ unless otherwise noted.

\subsection*{Definition of Transfer Entropy}

Transfer entropy quantifies the improved predictability about the future of a given temporal sequence $Y$ when including the history of another temporal sequence $X$, over just the history of $Y$.  Thereby, it attempts to measure the influence of $X$ on $Y$, and in the context of neuronal networks, measure the functional connectivity from $X$ to $Y$.
Specifically, the transfer entropy from the discrete sequence of random variables $X_t$ to $Y_t$ is  expressed as \cite{te1}
\begin{equation}
\label {TEeqn}
 TE_{X \to Y} = H(Y_t | Y^-) - H(Y_t | Y^-, X^-) 
\end{equation}
where $H(A | B)$ denotes the conditional Shannon entropy
\[ H(A|B) = \sum_{i,j} p_{A,B}(a_i,b_j) \log\frac{p_{B}(b_j)}{p_{A,B}(a_i,b_j)}. \]
Here, we write the formula for discrete random variables $A$ and $B$ with probability mass functions $p_A(a)$, $p_B(b)$ and joint probability mass function $p_{A,B}(a,b)$.  In Eq (\ref{TEeqn}), $Y_t$ is the value of sequence $Y$ at time $t$, and $X^-$ and $Y^-$ refer to the past states of the sequences $X_t$ and $Y_t$.

In practice, we do not know the true distributions of the random variables, only a collection of observations. The probability distributions must therefore be estimated from these observations.
We use a MATLAB toolbox~\cite{tet}
to numerically estimate the discrete transfer entropy values. 
The code takes as input a collection of sequences of sorted spike times. For each pair, it counts occurrences of spikes in both processes separated by a specified delay, and uses these counts to estimate the probability distributions that appear in the TE formula, resulting in a TE value for each pair of processes. This toolbox takes parameters for the interaction delay, and the history length for the source and target neurons.  In this paper, we use a history length of 5 for the source neuron and 3 for the target neuron.
To determine the interaction delay, we compute TE values for a range of possible delays and for each pair independently take the maximum TE value over this range.
For the sake of computational efficiency, we consider delays that are multiples of 10 time-bins  (0.01), up to a maximum of 200 time-bins (0.2).

Note that with this estimation method, even totally independent sequences will likely produce positive transfer entropy values due to statistical noise. This means that we can only claim that a positive transfer entropy value indicates information transfer 
 if it is large relative to the bias of the estimator, which depends on the input series. There is further discussion of this problem in Properties of Transfer Entropy on our Model section in Results.

Throughout the paper we only consider detecting information transfer when the source neuron is excitatory.  Although an inhibitory connection will theoretically produce a non-zero TE value, for relatively infrequent firing regimes, the absence of a firing event contains much less information than the presence of a firing event induced by an excitatory connection.  Here, infrequent firing is relative to the time-binning of the data;  a small fraction of time bins contain a firing event.

\subsection*{Distribution-based Comparisons}

In order to automatically handle the effects of varying neuron firing rate and other factors on the estimated TE values, we consider the usage of pairwise methods that determine the significance of the TE value between each pair of neurons in isolation from all other pairs. This prevents the possibility of neurons being overshadowed by other neurons with higher estimated TE values.

Prior literature has used methods in which a statistical baseline is computed using some sort of Monte Carlo method, and then the TE values are compared to this baseline distribution~\cite{bauer2007finding}.  Using a baseline distribution can account for fluctuations in properties like firing rate as long as the assumption that each observation comes from the same underlying distribution.
Inspired by typical experimental data consisting of multiple repeated trials, we create a baseline for each pair of neurons by computing the TE from the source neuron in one trial to the target neuron in a different trial.  Due to the time separation between trials, there cannot be actual transfer of information between the neurons, and any positive TE values must therefore be insignificant.  The signed-rank test is used to determine if  the distribution of estimated TE values from each of the trials is significantly greater than the distribution of TE values taken from time-shuffled comparisons.  We next detail the trial-shuffle method, other methods for creating this baseline, the false-positive method, and then describe how we determine overall connectivity between regions.

\subsubsection*{Trial Shuffle}

The primary method considered in this paper is the trial-shuffle method, in which 
 a number of ``replaced" trials are assembled, where the data for each neuron is taken from a different original trial. For example, one might combine the data for neuron 1 from trial 1, neuron 2 from trial 2, etc. into a single replaced trial.  By assembling the data this way, we can compute all the pairwise TE values for a shuffled trial and produce baseline values for all pairs simultaneously.  Since typically fewer neurons are observed than trials exist, it is possible to construct data in this way, with each spike train in a shuffled trial coming from a different experimental trial.  
This method is also not limited to having fewer observed neurons than trials.  In such a case, not all the baselines could be computed simultaneously, but could still be computed independently.

From the shuffled trials, we obtain a distribution of insignificant TE for each individual pair of neurons to serve as the baseline distributions for comparison. At the same time, we compute all of the pairwise TE values for each of the experimental trials, without shuffling. These values are collected to create the experimental distribution of TE values for each pair of neurons. At this point, we deem the connection to exhibit a significant transfer of information if the values in the experimental distribution for that pair of neurons are statistically significantly greater than those in the respective baseline distribution.

\subsubsection*{ISI Shuffle}
For the ISI shuffle~\cite{rivlin2006local}, the baseline spike trains are created by shuffling the relative offsets between spike times. In other words, if spikes occur at times $t_1,t_2,t_3,...$ and have offsets between them of $o_1 = t_1, o_2 = t_2-t_1,o_3 = t_3-t_2,...$ we shuffle the set of $\{o_i\}$ and reconstruct the spike times accordingly.

\subsubsection*{Jittering}
For jittering~\cite{nigam2016rich}, surrogate baseline data is produced by moving each spike time in an existing spike train by a certain random offset. In this paper, we generate the offsets using a uniform random distribution from $-20$ time-bins to $20$ time-bins, with resampling if the new location bin already contains a spike.

\subsubsection*{Time Shuffle}

For the time shuffle, a pairwise baseline is created by randomizing the spike times of each trial, such that each spike time is moved to a new time that is uniformly randomly chosen within the total range of time, with resampling if shuffled spike times fall in the same time-bin.

\subsubsection*{False Positive Method}

 The false positive method~\cite{tet}, unlike the other methods presented above, is not a pairwise comparison.  In \cite{tet}, all computed TE values are compared to a single cutoff chosen to enforce a certain false positive rate.
To define a false positive, we classify a directed pair of neurons $(a,b)$ as such:
\benum
\item
Direct Connection - Neuron $a$ has a direct structural connection to neuron $b$ ($c_{ab} \neq 0$)
\item
Indirect Connection - A sequence of multiple structural connections connect neuron $a$ to neuron $b$.
\item
No Connection - It is impossible to reach neuron $b$ from neuron $a$ traveling through the network.
\eenum
We define a false positive as a connection of type 3, where no transfer of information is possible, that is detected as having a significant transfer of information. 
To extend this approach to the multiple trial data format, we take the mean TE value across all of the trials for each pair of neurons, and then apply the standard false positive cutoff approach by choosing the cutoff to be the value such that only 5\% of unconnected pairs of neurons have a mean TE value above the cutoff. A pair of neurons whose mean TE is above the cutoff is then deemed significant.

\subsubsection*{Assessing inter-region connectivity}
We are interested in being able to determine, from the pairwise connection results, whether there is evidence of information flow between the two regions in our model.  We test the number of significant connections between regions against the 
 null hypothesis that all pairwise connections have an independent 5\% chance of being detected.  This significance level is chosen because our tests are done at a 5\% significance level.  The 95th percentile of the binomial distribution $\text{Binomial}(n_{\text{obs}}^2/4,0.05)$ when observing 50 neurons,  $n_{obs}=50$, gives the critical value of $40$ connections.

\section*{Results}

\subsection*{Properties of Transfer Entropy on our Model}
\label {modelproperties}

We will begin by illustrating how the transfer entropy estimation is affected by the input. We show that the estimated TE values can drastically vary independent of the actual causal relation between the neurons. This demonstrates the need for careful determination of the significance of TE values to infer functional connectivity.

\subsubsection*{Relation between Transfer Entropy bias and Trajectory Length}
\label{lengthbias}

While theoretically, only correlated sequences would have non-zero transfer entropy values, in practice, the estimation of distributions leads to a positive bias in the estimated TE values.  By the law of large numbers, the bias decreases as the observation length of the processes increases and more data is used to estimate the distributions.  

To get an intuitive idea of how this error should behave, consider first the simpler case of trying to estimate entropy of a stationary process. Consider a stationary process $Y_t \in \{0,1\}$ consisting of a sequence of independent, identically distributed Bernoulli random variables with mean $p$. Clearly the entropy of such a process will be 
\[ H(Y) = -\left(p \log p + (1-p) \log(1-p)\right) = \th(p).\]

If we do not know $p$, and only know observations of $Y_t$ for $t=1,2,...m$, then we can't compute $\th(p)$. Instead, we can let $\hat{p}$ be an estimator of $p$ and then estimate the entropy as $\hat{H}(Y) = \th(\hat{p})$. The question arises: what will be the error in this estimate?

\newcommand{\Ep}{\mathbb{E}_{\hat{p}}}

Clearly $\th(p)$ is a concave function, which tells us that for any imperfect estimator $\hat{p}$, 
\[ \mathbb{E}_{\hat{p}} \theta(\hat{p}) < \theta(\mathbb{E}_{\hat{p}} \hat{p}) \]
by Jensen's inequality. We will specifically consider the estimator $\displaystyle{\hat{p}=\frac{1}{m}\sum_{t=1}^m Y_t }$ which is unbiased ($\mathbb{E}_{\hat{p}} \hat{p}=p$) so from the above equation it will always produce an entropy estimate less than the true value. The estimator $\hat{p}$ is distributed as $\frac{1}{m}\text{Binomial}(m,p)$, and by taking a Taylor expansion about the peak $p$ of the distribution of $\hat{p}$,
\[ \Ep \th(\hat{p}) \approx \th(p) + \th'(p) \Ep[\hat{p}-p] + \frac{1}{2} \th''(p) \Ep[(\hat{p}-p)^2]  = \th(p) + \frac{\th''(p)}{2} \text{Var}(\hat{p}) . \]

We find that the bias in this entropy estimator is approximately
\[ \frac{1}{2} \th''(p) \text{Var}(\hat{p})  = \frac{1}{2} \left(-\frac{1}{1-p} - \frac{1}{p}\right) \frac{p(1-p)}{n} = -\frac{1}{2n}. \]

\begin{figure}[h!]
\centering
\includegraphics[width=0.7\textwidth]{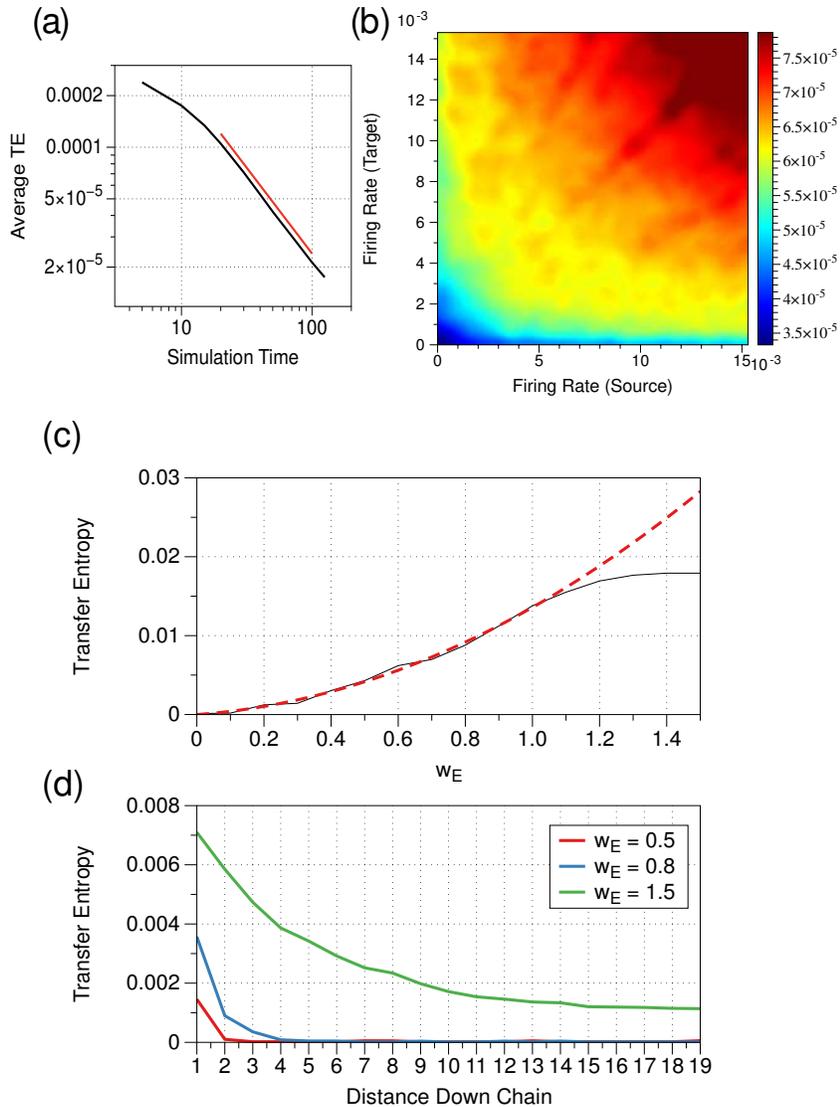}
\caption{(a) Black line:  Relationship between observation length and estimated TE averaged over a simulated network of disconnected neurons. Red line: sample slope of $-1$. For this simulation $n=50$ and $\mu=1$. (b) Relationship between neuron firing rate and calculated TE values showing an asymmetric non-linear dependence. The simulation is performed with $n=200$ unconnected neurons. $\mu$ values range linearly from 0 to 16, and $t_{max} = 10$. (c) Black line: Relationship between coupling strength and TE values showing an initial quadratic dependence before saturating due to the target neuron always firing after the source neuron. Red dashed line: Quadratic model fit to data for $w_E$ from $0$ to $1$, with  $R^2 = 0.99552$ (fit parameters in Appendix A).  For this simulation, $n=2$, $t_{max}=10$,$\mu=0.25$ and $w_E$ is given on the \textit{x}-axis. (d) Computed TE values for indirectly connected neurons down a chain with three different coupling strengths $w_E = 0.5,0.8,1.5$ showing that higher $w_E$ causes significant TE values further down the chain.  For this simulation $n=20$, $\mu=0.8$, $t_{max} = 100$.}
\label{fig:properties}
\end{figure}

Given this inverse relationship with the sample length in the simple entropy case, we might expect to see the same inverse relationship when estimating transfer entropy of our spike-train data.
To demonstrate this relationship, we 
 simulated a network of 50 neurons with no connections, and estimated the TE for various observation lengths. These TE values, averaged over the population, are shown in Fig~\ref{fig:properties}(a). We can see that the estimated TE between independent processes does decay with increased observation time, and for sufficiently long time it follows an inverse relationship.

\subsubsection*{Firing Rate vs. Transfer Entropy}
\label{firingrate}

When determining significance, one might attempt to compare computed TE values across all pairs in the network to obtain a 
 single cutoff value, above which indicates a transfer of information. However, since factors like firing rate affect the magnitude of the TE values, pairs with greater TE values do not necessarily have a stronger flow of information or even a connection at all.  Intuitively, this increase in TE value with firing rate can be understood by considering that spike-train data  consists of a sequence of correlated Bernoulli random variables which maximize their entropy at $p=0.5$. Because the neuron firing rate per bin is much less than $0.5$, an increase in firing rate increases the entropy of the process and correspondingly increases the upper bound for the transfer entropy.  Due to the nature of the estimator, we would also expect that independent processes have higher estimated TE values when firing rates are higher.

Given this, we would like to understand the effect that the firing rate of neurons has on the estimated TE between them. We expect that higher firing rate neurons will have inflated TE values, independent of whether or not the pairs are actually connected. Fig~\ref{fig:properties}(b) shows the sharp increase in TE values for higher firing rate neurons, for a collection of neurons with no connections between them. This demonstrates why some sort of normalization is required if one wishes to compare estimated TE values from different pairs of neurons with different firing rates, especially for a collection of heterogeneous neurons as one might find {\em{in vivo}}.  This normalization is difficult even for just the firing rate due to its asymmetric non-linear dependence shown in Fig.~\ref{fig:properties}(b).  Furthermore,  firing rate is not the only factor influencing TE values.

\subsubsection*{Coupling Strength vs. Transfer Entropy}

Naturally, neurons connected with a higher coupling strength will generally have higher TE values observed between them.  
This bears relevance on our ability to detect weaker connections, which are harder to detect when the signal-to-noise ratio is low.  We illustrate this effect with a pair of neurons by varying their (excitatory) coupling strength.

The data, as shown in Fig~\ref{fig:properties}(c), suggests that for small coupling strength the estimated TE values are roughly proportional to the square of the coupling strength. As might be expected, once the coupling strength is comparable to the distance between the reset value and the firing threshold of the model neurons, the target neuron always fires after the source neuron does and therefore further increasing the coupling strength does not actually increase the transfer of information from the source neuron to the target neuron, causing the TE value to plateau.   Notice how a low-firing rate, strongly coupled neuron pair and a high-firing rate, weakly coupled neuron pair could therefore both produce the same TE value.

\subsubsection*{Indirect Connections}

Another issue in detecting functional connections in experimental data is that 
typically only a fraction of neurons are observed, so it is unlikely
that one would always observe pairs of directly connected neurons. In order to detect functional connectivity, we must therefore be able to detect indirect connections. Here, we review how estimated transfer entropy values are affected by increasingly indirect connections.
Fig~\ref{fig:properties}(d) shows the transfer entropy computed down the length of three chains of varying coupling strength. All three chains show that the direct connection from the head neuron to the first neuron down the chain is very significant, but they vary significantly in how the transfer entropy values decay further down the chain. In the weakest coupling strength chain, the values decay almost immediately to noise levels, meaning detecting more than one or two neurons down the chain would be difficult. In the medium coupling strength chain, there appears to be an exponential decay, falling to noise levels after about 5 neurons. However, in the strongest coupling strength chain, the transfer entropy remains a significant fraction of the directly-connected neurons' TE value for the entire length of the chain.  This is not entirely surprising as  the large value of $w_E = 1.5$, relative to $V_T-V_R=1$ (the distance from reset to threshold voltage), allows for a wave of firing to frequently move from the head neuron all the way down the chain.  This wave of firing is only interrupted if the noise, which here has strength $\beta=1$, has driven the voltage of a neuron below -0.5 when the previous neuron fires.  Given this is fairly unlikely to happen, high TE values arise for even the furthest neuron down the chain.

\subsection*{Validation of Method}

We establish the validity of using the trial-shuffle method to further investigate flow of information by making a qualitative comparison to the other methods presented in Methods.  We begin by creating synthetic spike-train data according to the two-region network model as explained previously in Methods.  We simulate 64 trials and use this same data for each of the significance testing methods to identify connections.  
To parallel previous work~\cite{tet} we will report the fraction of connections that were identified for various lengths of connection, and we will also compare the number of false positives from each method.

\begin{figure}[h!]
\centering
\includegraphics[width=1\linewidth]{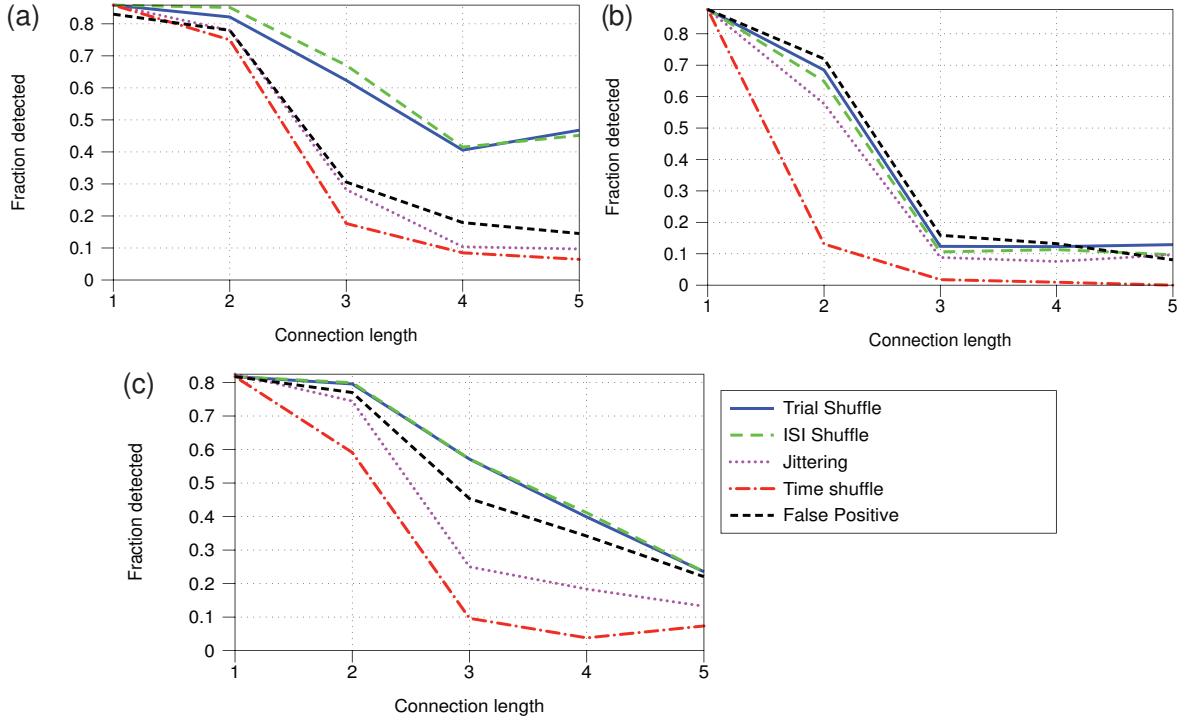}
\caption{Comparison of the false positive and distribution-based methods in terms of fraction of detected connections vs.~path length in the network.  For all frames $n=k_A=50,k_B = 0$, $t_{max}=10$. (a) $w_E=0.4$, $w_I=-0.5$, $\mu=10$  and $k_1=2$ (b) $w_E=0.27$, $w_I=-0.4$, $\mu=3$  and $k_1=2$   (c) $w_E=0.35$, $w_I=-0.45$, $\mu=2$  and $k_1=3$}
\label{fig:validation}
\end{figure}

\begin{figure}[h!]
\centering
\includegraphics[width=1\linewidth]{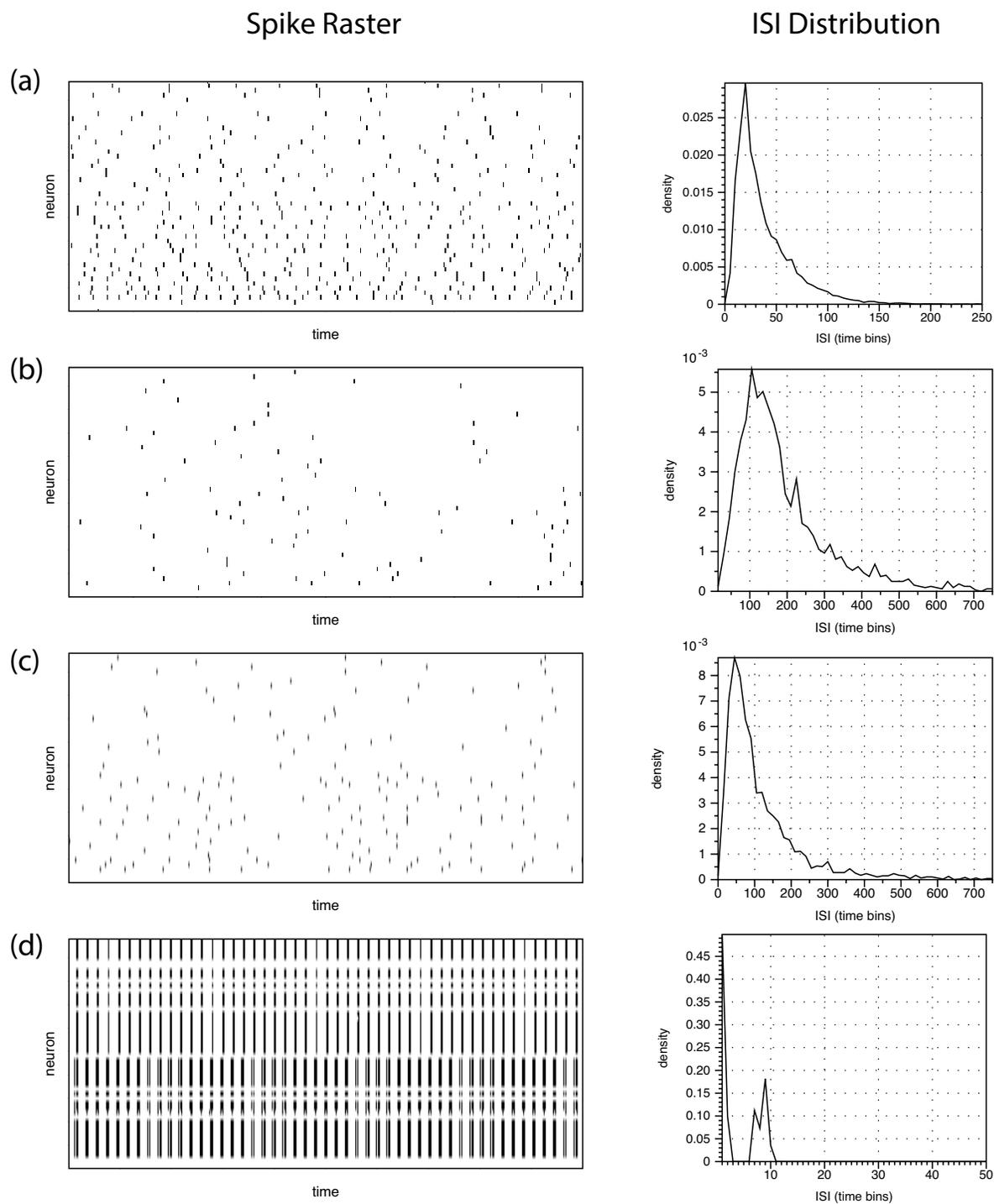}
\caption{Raster plots and inter-spike-interval (ISI) distribution for the two-region network model in various parameters regimes.  (a) through (c) correspond to (a) through (c) in Fig.~\ref{fig:validation}.  We try to avoid the dynamics shown in panel (d) with $k_1=10, w_E=0.5, w_I=-2, \mu=1$.  This case has no interesting transfer of information due to the highly synchronous behavior.}
\label{fig:raster}
\end{figure}

Fig~\ref{fig:validation} shows the percentage of connections present in the network that are identified as significant by each method for three different sets of model parameters.  
Recall, we only consider detecting connections starting from excitatory neurons.  Connections of length above 5 are not shown as they are not present in great enough quantity to produce meaningful information.  The network behavior of the neurons for each set of parameters tested in Fig.~\ref{fig:validation} is shown in Fig.~\ref{fig:raster}.

\newpage

While we are primarily interested in the four pairwise distribution comparison methods, we also show the false positive method for comparison. We reiterate that it requires as input the full network topology, which includes the true solution to the detection problem, and so is not a valid comparison. However, the parameters and randomly initialized structure of the network significantly affect how easy it is to detect connections in the network, and so the performance of the false positive method
provides a simple reference point for how easy it is to detect connections in a given network. We stress that comparisons should not be made between the false positive method and the other methods.

Throughout the parameter space (three typical cases shown) we see a general pattern of the trial-shuffle method and the ISI-shuffle method  detecting more connections, with the jittering method in the middle and the time-shuffle method  clearly detecting the least. These results indicate that  both the trial-shuffle method and the ISI-shuffle method would be reasonable options for further investigation of transfer entropy behavior.

Of course, the detection must also be judged in light of the false positive rate, as increased detection is not valuable if it is not restricted to those connections that are actually present. In Table~\ref{tab:falsepositive} we present the false positive rates associated with each of the four pairwise distribution comparison methods we consider. Recall that all statistical tests were performed at a 5\% significance level. 

\begin{table}[h]
\centering
\begin{tabular}{c|cc|cc|cc}

& \multicolumn{2}{c|}{panel a} & \multicolumn{2}{c|}{panel b}& \multicolumn{2}{c}{panel c} \\

method & all & B$\to$A &all & B$\to$A &all & B$\to$A  \\
\hline
Trial Shuffle & 7.7&4.2    &     3.7&2.5    &    3.6&2.5 \\
ISI Shuffle & 10.0&6.4 & 4.5&4.2 & 6.9&6.4 \\
Jittering & 4.1&1.7 & 4.9&4.1 & 3.8&2.4 \\
Time Shuffle & 1.4&0.16 & 2.2&1.6 & 5.4&4.9 
\end{tabular}
\caption{False positive rates corresponding to Fig~\ref{fig:validation}.  ``All'' refers to connections found relative to all pairs without any type of structural connection.  ``B$\to$A'' refers to just connections from region B to region A for which no connections of any kind exist.}
\label{tab:falsepositive}
\end{table}

We break up the false positive rates into the overall false positive rate and the $B \to A$ false positive rate. Recall that the network contains two regions, $A$ and $B$, for which inter-region connections exist only from $A$ to $B$. Because our research is motivated by the search for macroscale information flow, we are particularly interested in avoiding those false positives that indicate fictitious transfer of informations between regions, i.e.~$B\to A$.

We see that the trial-shuffle method has lower false positive rates than the ISI-shuffle method which had similar levels of detection. The jittering method has similar false positive rates.
The time-shuffle method has very low false positive rates in panels (a) and (b), which when combined with its low true detection rate indicate that this method of creating surrogate data leads to difficulty finding any connections at all.

From this data we can conclude that the trial-shuffle method does appear to be a reasonable way to create surrogate data to test for transfer entropy significance. We will now investigate how the model and simulation parameters affect the ability of the trial-shuffle method to find connections in a two-region network.

\subsection*{Inferring Information Transfer}
We will now show the detection of information flow as a function of network parameters coupling strength and connectivity.  Specifically, we will use the trial-shuffle method to determine global causal links between regions in the two-region network model.

\begin{figure}[h]
\centering
\includegraphics[width=0.9\textwidth]{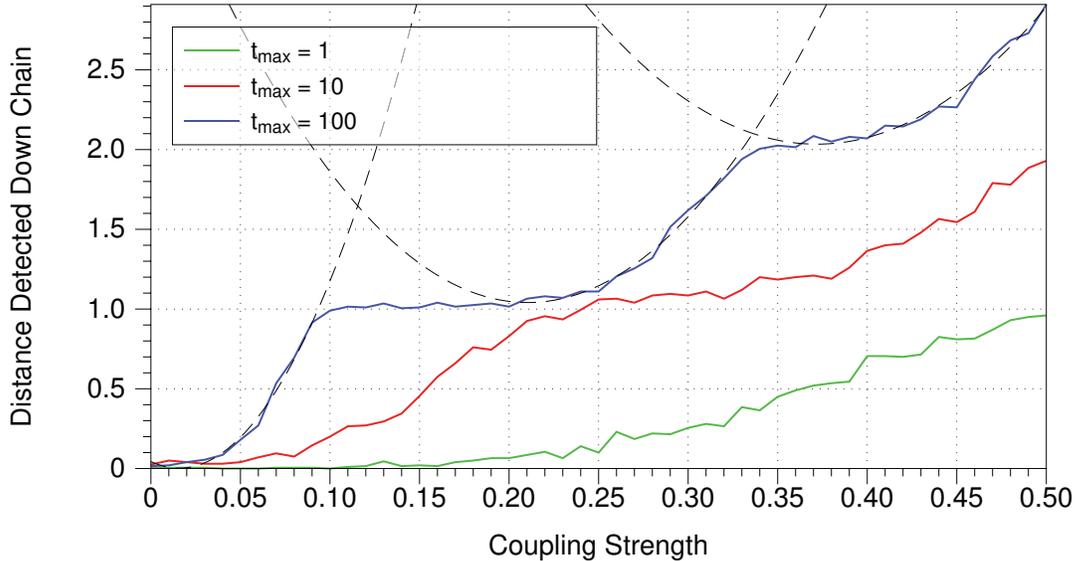}
\caption{For a range of connection strengths $w_E$ and three total simulation times $t_{max}$, how many neurons down the chain are detected as connected to the head, on average. 
The leftmost quadratic model is fit on $x$ from $0$ to $0.09$, with $R^2 = 0.99109$. The middle quadratic model is fit on $x$ from $0.2$ to $0.32$, with $R^2 = 0.98718$. The rightmost quadratic model is fit on $x$ from $0.35$ to $0.5$, with  $R^2 = 0.98325$. (fit parameters in Appendix A)
 For this figure we use $n=20$, $\mu=1$. All connections are excitatory with coupling strength as given on the \textit{x}-axis, and $t_{max}$ is specified in the legend.}
\label{wEvsDetection}
\end{figure}

A natural question is the relationship between the coupling strength of two neurons and our ability to detect a connection between them. Obviously, we would expect a stronger excitatory connection to be more likely to be detected as significant. To investigate this, we created a  directed chain network of 20 neurons, and investigated the detected connections originating from the neuron at the head of the chain. In theory, this neuron is connected to all 19 neurons down the chain from it; however, when the coupling strength is low these connections will likely not all be able to be detected. For an ensemble of 1000 trials, we observed how many neurons down the chain were detected as connected to the head neuron, and these results are shown in Fig~\ref{wEvsDetection}. 
We therefore see, depending on the coupling strength, how long a sample must be taken to reliably detect direct connections or some indirect connections.

%
%
%
%

Especially visible in the $t_{max}=100$ case is the nonlinear structure of the data. As the coupling strength increases, the probability of detecting a connection to the first neuron increases quadratically. This intuitively aligns with what we previously found in Fig~\ref{fig:properties}(c), that the transfer entropy itself increases quadratically with coupling strength. After the coupling strength is high enough that the first connection can be very consistently detected, the second connection does not immediately become detectable. Instead, there is a region in which increasing the coupling strength does not increase the length of connection detected. 
 
The most important thing we want to understand is how the degree of connectivity in the two-region network relates to the ability of our method to correctly identify the flow of information. We will be again considering the two-region network with monodirectional flow, e.g. from region $A \to B$ but not from $B \to A$. The connectivity can be influenced both by the edge probability parameters $k_1$ and $k_A$ and the number of neurons $N$ in the network. We would like to find a way to quantify the amount of inter-region connectivity across networks with variation in all of these parameters. To this end, we introduce a variation of the global efficiency~\cite{latora2001efficient}, which is the average of the inverse shortest path length between all pairs of nodes. Our variation measures the directed efficiency from region A to B by considering only those paths from a neuron in region A to a neuron in region B, and then averaging the inverse shortest path length.

\begin{figure}[h]
\centering
\includegraphics[width=0.9\textwidth]{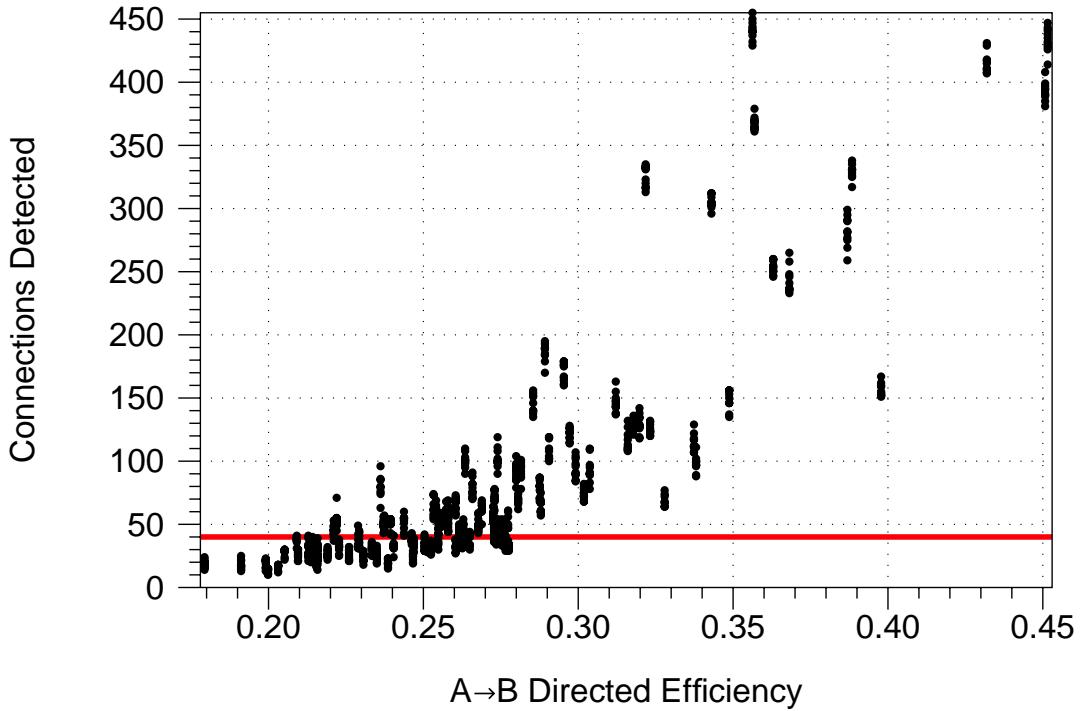}
\caption{$A \to B$ efficiency vs detection. The red line indicates the significant level of connections indicating transfer of information between regions. Here we use $t_{max}=10$, $\mu=1$, $w_E = 0.5$, $w_I = -1$. The networks are randomly initialized with $n$ ranging from $50$ to $1000$ with $n_{\text{obs}} = 50$, $k_1 \in \{2,3\}$, and $k_A$ ranging from $\frac{n}{2}$ to $2n$ with $k_B = 0$.}
\label{k2d1}
\end{figure}

 We generated a large number of networks of varying sizes from $n=50$ to $1000$, only observe $n_{\textrm{obs}}=50$ neurons, $k_1 =2 \text{ or } 3$, and $k_A \in \{n/2,n/2+1,\dots,2n\}$, leading to a wide range of directed efficiencies. For each of these networks we run 10 sessions of 64 trials to get a quantification of the spread in potential detection. Fig~\ref{k2d1} shows the number of detected connections for each network against the directed efficiency. This confirms the expected, that more connected networks have a higher number of connections detected. At an efficiency of approximately $0.28$, the method begins to always detect the regions as being connected, although it frequently still does so for lower values. Also of interest is the fairly narrow spread in detection across the substantial variation in parameters for the lower efficiency range. This suggests that our method is stable across a variety of networks.
 
\begin{figure}[h]
\centering
\includegraphics[width=0.9\textwidth]{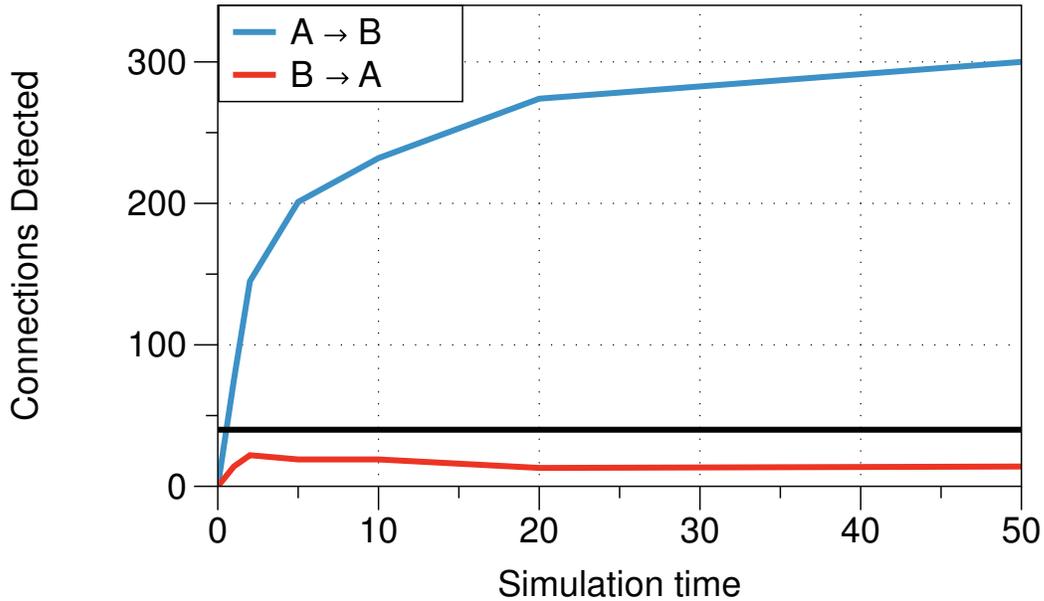}
\caption{Number of detected connections vs. simulation time for each trial. Blue = $A\to B$, red = $B \to A$. The black line indicates the level above which the number of connections is statistically significant ($p<0.05$). This simulation uses a two-region network with $n=50$, $k_1=3$, $k_A=50$,$\mu=3$,$w_E=0.27$,$w_I = -0.3$, and $t_{max}$ given on the \textit{x}-axis.}
\label{time1}
\end{figure}

We would also like to see how the ability of our method to detect connections improves using trials of increasing time length. Recall Fig~\ref{fig:properties}(a), which shows that the noise in the TE estimator is inversely proportional to the length of time in the trial. This suggests that a longer trial should have a higher signal-to-noise ratio and correspondingly better detection. Additionally, since neurons are actually connected in this network, the longer time means that there are more chances for a neuron to affect another, meaning there is more information transfer for the method to find. Fig~\ref{time1} shows the number of significant connections detected between regions in each direction, for a network that only contained $A \to B$ inter-region connections with 50 neurons. 
In line with our expectations, we see a steady increase in the number of connections we detect as the length of the simulation increases. 

\section*{Discussion}

We use a pairwise comparison method to determine the significance of transfer entropy values in an objective way for spike-train data comprised of multiple trials.
By shuffling trials, we create surrogate data to serve as a baseline for insignificant TE values and then performing a statistical test to see if the true TE values are greater than the baseline TE values. We also compare to several other options for creating such surrogate data: shuffling the spike times, shuffling the inter-spike-intervals, and jittering the spike times. Our analysis showed that the trial-shuffle and ISI-shuffle methods had the highest detection, and the trial-shuffle method had slightly lower false positive rates, which justified our decision to look further into the behavior specifically of the trial-shuffle method.
We hypothesize that the reason why the trial-shuffle and ISI-shuffle methods have the highest detection is that while all of the methods preserve the firing rate in the spike trains, only the trial shuffle and ISI shuffle preserve the temporal structure of the data.

This trial shuffling method focused on data separated into trials, to allow for a coarse-grained shuffling in time. The time shuffling method is another extreme of this shuffling in which the data is shuffled on the order of a single observation time step. The fact that the time-shuffle method fails to accurately determine significance whereas the trial-shuffle method does not suggests that shuffling is only valid when done on a sufficiently coarse timescale. There is presumably some ``minimum time window" that could be shuffled, such that the temporal structure of the spike trains is still preserved. We hypothesize that this time window must be greater than the timescale at which information flows through the network. Breaking up data on such a timescale and then shuffling it will allow the trial-shuffle method to be extended to spike-train data consisting only of a single long recording, instead of a number of separate recorded trials, as long as analysis confirmed a robust choice of shuffling timescale.

We observed interesting features of information flow on a network.  Looking at how far down a directed chain of connected neurons transfer entropy detects significant information flow revealed a quadratic dependence of the distance down the chain on the coupling strength.  This aligns with the observed quadratic dependence of the estimated transfer entropy on the coupling strength.  We also observed the global flow of information from one network region to the other as a function of network connectivity.  While increased detection of connections is expected as the connectivity of the network increases, we note the fairly narrow spread in detection across substantial variation in parameters.  One might expect these general trends to remain when considering different transfer entropy estimating methods, or more realistic models for producing spike-train data.

A modification to the neuronal network model could allow for testing on a more organized network in future work. Various methods exist to organize information flow in a neuronal network, including Spike-Timing Dependent Plasticity (STDP) and other extensions of the Hebbian learning philosophy. A meaningful external drive could also be applied to create a source of non-random information. These methods could be used to create a network that has more concentrated connections that direct information flow instead of diffusing it as in a random network. Computational advances could also help bridge the wide gap between the number of neurons we used in simulated networks, up to 1000, and the many billions of neurons actually present in the human brain. This would allow us to see the effect that the vastly sparser observation has on the results, when coupled with the increase in organization as described above.

In this paper we chose to use a basic model for simulating spike-train data. However, more sophisticated models such as the Izhikevich model~\cite{Izhikevich2003simple} or a Hodgkin-Huxley model with adaptation~\cite{woo2009simulation} could lead to different sorts of spiking behaviors in addition to greater historical dependence of the current behavior. These effects would certainly affect the problem of detecting TE values, and so would provide an interesting followup to the work presented in this paper.

 While we chose one transfer entropy estimator in our analysis, we expect many of the features we observed in the behavior of information detection to remain regardless of the TE estimator.  Many variations and extensions of transfer entropy have been explored in the literature.  For example, Ito et al.~\cite{tet} suggest that first order TE might be better than higher order TE unless the orders are tuned.  Multivariate extensions of transfer entropy, which in addition to controlling for information contained in the past of the target process, can control for information present in other processes besides the source, provide transfer entropy values that reflect the information coming from the source neuron and not from any of the other neurons. 
These tools, combined with an effective method to determine significance such as the trial-shuffle method, could be used to create functional connectivity graphs that show how information is actually moving between the neurons based on spike-train data. This would then allow a more in-depth understanding of how the neuronal networks in the brain actually processes information.

Real-world data brings many additional challenges when looking for functional connectivity. For one, the trial-shuffle method assumes each trial is an independent sample from the same distribution, thus cannot account for adaptation or depression across multiple trials.  Also, we expect connections to be in general weaker, and the number of neurons is many times greater than used in our simulated tests. This is balanced to a degree by the fact that neurons in the brain are extremely well organized to facilitate the flow of information, in contrast to the random connections in our simulated network. This may lead to greater sensitivity of the method, as detection could depend heavily on whether an organized pathway is actually picked up by the probes.

We might try to mitigate the real-world challenge of detection by obtaining more data, as we have shown the ability to detect connections scales with the amount of observed data. Depending on the nature of the experiment, it may or may not be possible to increase the length of the observed trials. Instead, more data could be obtained by increasing the number of trials. However, the analysis requires that the same neurons be recorded consistently over all of the trials, and small movements of the probes over time can potentially cause neurons to be lost in the recording. These challenges in obtaining large amounts of experimental data highlight the importance of a statistical analysis that is as powerful as possible.

\appendix

\section{Quadratic Fit Parameters}

The parabola fit in Fig.~\ref{fig:properties}(c) is $y=-8.6573 \times 10^{-6} + 0.0030409 x + 0.010546 x^2$.

\noindent The parabola fits in Fig.~\ref{wEvsDetection} are $y = 0.043545  -5.2788x + 166.67 x^2$ for the leftmost, $y=4.0181 -28.291x + 67.208x^2$ for the middle, and $y = 9.2834 -38.99x + 52.425x^2$ for the rightmost.

%
%
%
%


\end{document}